%% file: frontiers.tex
\documentclass[utf8]{FrontiersinHarvard} 

\usepackage{url,hyperref,microtype,subcaption}
\usepackage[onehalfspacing]{setspace}
\usepackage{comment}

\def\sys{\texttt{sys2025}}

\def\keyFont{\fontsize{8}{11}\helveticabold }
\def\firstAuthorLast{Bonamente} 
\def\Authors{Massimiliano Bonamente\,$^{1,*}$}
\def\Address{$^{1}$Department of Physics and Astronomy, University of Alabama in Huntsville, Huntsville, AL 35899 }
\def\corrAuthor{Corresponding Author}

\def\corrEmail{bonamem@uah.edu}

\input{defs}

\begin{document}
\onecolumn

\title[Systematics in Poisson regression]{Review: A new method for estimation and use of systematic errors in Poisson regression} 

\author[\firstAuthorLast ]{\Authors} 
\address{} 
\correspondance{} 

\extraAuth{}

\maketitle

\begin{abstract}


A new method for including systematic errors in the regression with Poisson data is reviewed
in this contribution, with emphasis on applications to astronomical spectra. The method consists
of generalizing the usual Poisson log-likelihood, known as the Cash statistic \cmin, and its associated likelihood-ratio statistic
\DeltaC, to include the presence of systematics sources of uncertainty. Advantages of this new
method include its modeling simplicity and its ability to assess both the level of 
systematics and the goodness of fit at the same time, including for a nested model component.

\tiny
 \keyFont{ \section{Keywords:} Poisson regression, systematic errors, goodness of fit, Cash statistic} 
\end{abstract}

\section{Introduction}
In the physical sciences and related disciplines 
uncertainties or `errors' in quantities of interest are often
categorized into two types:
`statistical' errors associated with the method of measurement, and `systematic' errors that are due to unknown sources of uncertainty, either in the measurement 
process itself or in the underlying theory. 
A review by \cite{vandyk2023} describes several methods that are commonly used in the physical sciences
to address sources of systematic error.

For Poisson count data, the fixed relationship between the mean and variance of the distribution complicates the statistical treatment of systematic errors, compared to
the case of Gaussian statistics.
Overdispersed count--data distributions, such as the negative binomial 
or the Poisson inverse Gaussian are a possible alternative \citep[see, e.g.,][]{hilbe2011,hilbe2014, cameron2013}. 
In \cite{bonamente2025a} and \cite{bonamente2025b} we have presented a new method
for Poisson regression and associated methods of goodness of fit assessment that we review
in this contribution. The goal of this method is to retain the Poisson distribution for the data,
and to add a source of systematic error associated with the model itself. This method leads to a
simple extension of the usual Poisson regression that can be conveniently applied in 
virtually all data analysis situations. Key features of this model are reviewed in this contribution,
and illustrated with an example with astronomical data.

\section{Review of Methods} 
\label{sec:review}
In this section we review the main method of analysis for the regression of Poisson data 
in the presence of systematic errors in the models presented in \cite{bonamente2025a, bonamente2025b}. It is assumed that the data are of the
form of $N$ independent measurements 
\begin{equation}
y_i \sim \Poiss(M_i)    
\label{eq:yi}
\end{equation}
 made at a fixed value $x_i$
of an independent variable that typically represents energy or wavelength for a spectrum,
or time for a light curve. 

In the absence of systematic errors, one would normally take $M_i = \mu_i$ to be a
\emph{parent} (i.e., fixed) mean that depends on a few adjustable parameters, $\mu_i(\theta)$, with $\theta$
being $m$ adjustable parameters. For example, the linear model would be $\mu_i(\theta)=\theta_1+x_i \cdot\theta_2$ with $m=2$ two parameters $(\theta_1,\theta_2)$ to be estimated from the data.
Under this baseline model, one uses the Poisson log-likelihood, known in the
astronomical community as the \emph{Cash} statistic \citep[e.g.][]{cash1979,baker1984},
to estimate the best-fit parameters $\hat{\theta}$. Goodness of fit is then
performed by means of the \cmin\ statistic, defined as
\begin{equation}
    \cmineq = 2 \sum\limits_{i=1}^N \left( y_i \ln \left(\dfrac{y_i}{\hat{\mu}_i}\right) - (y_i-\hat{\mu}_i) \right).
  \label{eq:cmin}
\end{equation}
The statistic is asymptotically chi-squared distributed in the limit of large means,
whereas the statistic approximates the usual $\chi^2_{\mathrm{min}}$ statistic
\citep[e.g.][]{bonamente2022book}. In general,
however, 
it has been shown that this statistic can be considered as asymptotically normal even in the
low-count regime, provided that the data are in the extensive limit of $N \gg 1$ . This is in fact a new result that had not been previously appreciated, and it is discussed in a separate paper \citep{li2026}. 

The main modification for the presence of systematic errors is to assume that the 
quantity $M_i$ in \eqref{eq:yi} is not fixed to a parent (i.e., true-yet-unknown value) 
$\mu_i$, but it is a \emph{random variable}. This variability is what introduces the
source of systematic errors by means of 
\begin{equation}
    M_i: \quad \text{random variable with} \quad \E[M_i]=\hat{\mu}_i,\; \Var(M_i)=\sigmaIntiEq^2.
    \label{eq:Mi}
\end{equation}
In \eqref{eq:Mi}, there are two main features to highlight: (a) the distribution of the random
variable can be any positive-valued random variable, since the mean of a Poisson distribution
cannot be negative~\footnote{Regardless of this limitation, one can in fact assume $M_i$ to be normally distributed in many applications by ignoring the left-tails  of its distribution, as discussed in \cite{bonamente2025a}}; and (b) the newly-introduced $\sigmaIntiEq$ parameter
represents the degree of systematic error or intrinsic variability of the model, \emph{after} the
usual Poisson regression is performed to obtain the usual $\hat{\mu}_i$ estimates. In practice it is convenient to parameterize this  term
as

 \begin{equation}
    f_i = \dfrac{\sigmaIntiEq}{\hat{\mu}_i} \ll 1 \quad \text{(relative systematic error).}
    \label{eq:f}
\end{equation}   
Usually the $\sigmaIntiEq$ parameter is taken to be constant among all data points,  meaning that a constant (absolute) systematic error
is assumed (and
the $i$
index becomes unnecessary).  But a different parameterization that allows a bin-dependent relative systematic error is also possible within this framework, as
discussed in \cite{bonamente2025a}. The relative error is intended to be a small correction and therefore
at the few percent level (typically $\leq$ 10~\%), for this model to be practical. Fortunately, this is the usual
level of systematics in most data for astronomy and the physical sciences. Larger systematic
errors would require an overall rethinking of the fit function, and they cannot be treated by this method.

With this model of systematic errors, the goodness-of-fit statistic is modified from \eqref{eq:cmin} to become
\begin{equation}
\begin{aligned}
    \cminsyseq = 
    &
     2 \sum_{i=1}^N \left( y_i \ln \left(\dfrac{y_i}{M_i}\right) - (y_i-M_i) \right) = X+Y.
    \end{aligned}
  \label{eq:cminMi}
\end{equation}
as the sum of two \emph{independent} contributions, the first of which is the usual
\cmin\ statistic of \eqref{eq:cmin}, and the second represents the additional
contribution from systematic errors.  It is necessary to point out that, as also noted after Eq.~\eqref{eq:Mi}, the additional variance and therefore the $Y$ contribution to \cminsys\ is introduced only after the usual Poisson fit with \cmin\ is made. This is by design, and it means that the best-fit model is the same whether \cmin\ or \cminsys\ is used for goodness-of-fit estimation.

It is also found that this additional contribution is approximately normally distributed as
\begin{equation}
    Y \overset{a}{\sim} N({\mu}_C,{\sigma}^2_{C}),
    \label{eq:YDistr}
\end{equation}
where the two parameters are respectively a \emph{bias} and an \emph{overdispersion} term.
It is therefore possible to do hypothesis testing with \cminsys, since its null-hypothesis
distribution can be calculated as the convolution of the two distributions for $X = \cmineq$
and $Y$ according to \eqref{eq:YDistr}.  For extensive data in the large-mean regime
(large value of $N$) both statistics are
approximately normally distributed, and therefore \cminsys\ continues to be normally distributed
with mean equal to the sum of the means, and variance equal to the sum of the variances.
In the low-mean limit, the mean and variance of \cmin\ can be calculated according to
a simplified method described in \cite{li2026}, and referred to as the \emph{KB approximations}
\citep{kaastra2017, bonamente2020}, where it is shown that the mean and variance can be substantially different from the asymptotic chi-squared values of 1 and 2 for each data points.
If one prefers to retain the chi-squared distribution for $X=\cmineq$, an analytic expression for the
convolution with a normal distribution is provided by the \emph{gamma-normal} family of distribution discussed in
\cite{bonamente2024properties}, also referred to as the \emph{overdispersed chi-squared} distribution.~\footnote{Recall that the chi-squared distribution is a special case of
a gamma distribution.}

There are two ways of using this model. If the level of systematic errors is known,
then $\sigmaIntiEq$ values are known (either constant or bin-dependent), and the two parameters can be estimated from the data  by
\begin{equation}
    \begin{cases}
         \hat{\mu}_C \simeq \sum_{i=1}^N  y_i\, f_i^2\\[10pt]
         \hat{\sigma}^2_{C} \simeq  4 \sum_{i=1}^N y_i\,  f_i^2, 
    \end{cases}
\end{equation}
whereas more accurate approximations are discussed in \cite{bonamente2025a}.

Alternatively, these parameters can be estimated from the data themselves, under the
assumption that the model is accurate. In this case, the bin-independent relative error is estimated
as
\begin{equation}
\widehat{f\,}=
\sqrt{\dfrac{\hat{\mu}_C}{\sum_{i=1}^N y_i }},\; \text{with }\; \hat{\mu}_C = \cminValEq -\E(X) >0
\label{eq:sigmaIntEst}
\end{equation}
and where \cminVal\ is the measured value of \cmin\ for the regression at hand. Clearly,
systematic errors can only be invoked when the measured value exceeds the expectation, i.e., when the
fit is poor. In this scenario, the main outcome of the method is the estimate of the 
level of systematics according to \eqref{eq:sigmaIntEst}, and hypothesis testing is not meaningful as one has already assumed
the applicability of the model.  It is worth noting that Eq.~\eqref{eq:sigmaIntEst} is a constant level of \emph{relative} systematic errors, which implies a non-uniform level of absolute estimated errors $\hat{\sigma}_{\mathrm{int},i}$. This distinction is often irrelevant for data model that do not vary significantly over the range of the independent variable, where
an overall estimate $\hat{f}$ is often sufficient. Alternatively,
one could parameterize the $\sigmaIntiEq$ values as a function of the independent variable $x$
and attempt an estimate of those parameters instead; this is an avenue for a possible
extension of this model of systematic errors.

This model of systematics is also extended to the \DeltaC\ statistic for the significance of
a nested model component, as described in \cite{bonamente2025b}. In the absence of
systematics, the statistic of interest is 
\begin{equation}
    \Delta C = C_{\mathrm{min,r}}-C_{\mathrm{min}},
    \label{eq:DeltaC}
\end{equation} 
with $C_{\mathrm{min,r}}$ being a reduced model where $k$ of the $m$ adjustable parameters
are held fixed at their true (null) values. The \DeltaC\ statistic is a likelihood-ratio
statistic, same as the popular $\Delta \chi^2$ statistics \citep[e.g.][]{cash1976,cash1979,avni1976}
that is commonly used for Gaussian data. It follows the usual Wilks theorem, so that
$\Delta C \sim \chi^2_{k}$ \citep[e.g.][]{wilks1938,wilks1962}, where $k$ is the number of additional parameters in the full model, relative to the reduced model. Usually we have $k=1$ or 2 for nested components that add respectively one or two additional parameters.

In \cite{bonamente2025b} we have shown that the above model for systematics leads to
the following modification:
\begin{equation}
    \DeltaCSysEq = (X_r-X) + (Y_r-Y) = \Delta X +  \Delta Y,
    \label{eq:DeltaCsys}
\end{equation}
where $\Delta X$ represents the usual \DeltaC\ statistic 
and the additional term $\Delta Y $ represents the contribution 
from systematic errors.  We found that $\Delta Y$ statistic has a Bessel distribution $K(\alpha)$, with $\alpha= 2 f \sqrt{\mu}$, i.e.,
it follows the density of a Bessel function of the second kind and of order zero, whereas
$\Delta X = \Delta C \sim \chi^2_{k}$. 

The null-hypothesis distribution of \DeltaCSys\ is somewhat complicated by the fact that $\Delta X$
and $\Delta Y$ may not be independent, which is a requirement for obtaining its distribution
from those of the two contributing statistics. In general, the convolution of a chi-squared and a
Bessel distribution is referred to as a \emph{randomized chi-squared} distribution \citep{bonamente2025b}. If the two contributions to \DeltaCSys\ are independent,
then an approximate analytic function for its distribution is found by approximating the
Bessel distribution via a Laplace distribution \citep{bonamente2025properties}, which leads to an analytic approximation of the {randomized chi-squared} distribution. Alternative, simple numerical methods can be used for determining the distribution of \DeltaCSys, including the
presence of correlation between the contributing terms, as explained in \cite{bonamente2025b}.



\section{Discussion}
\begin{figure}
    \centering
    \includegraphics[width=0.45\linewidth]{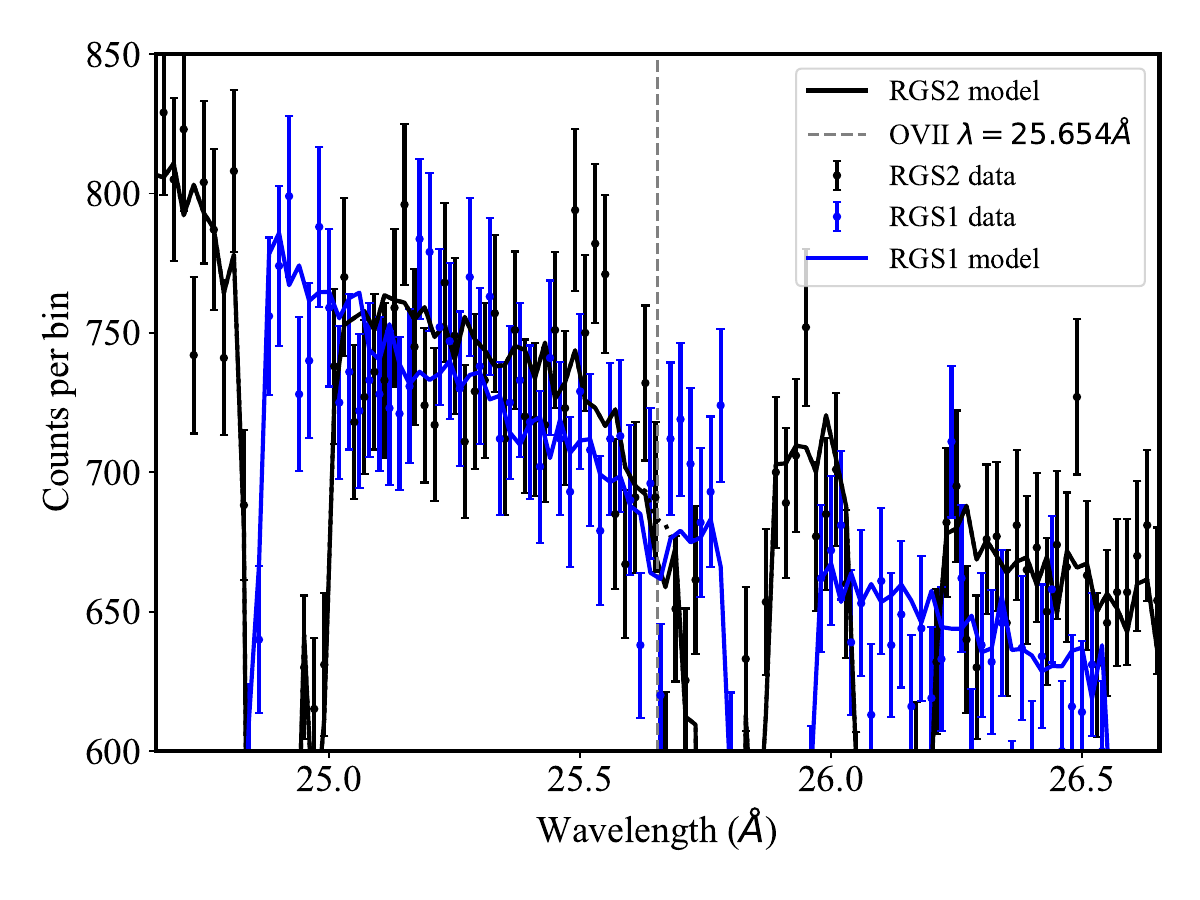}
    \includegraphics[width=0.45\linewidth]{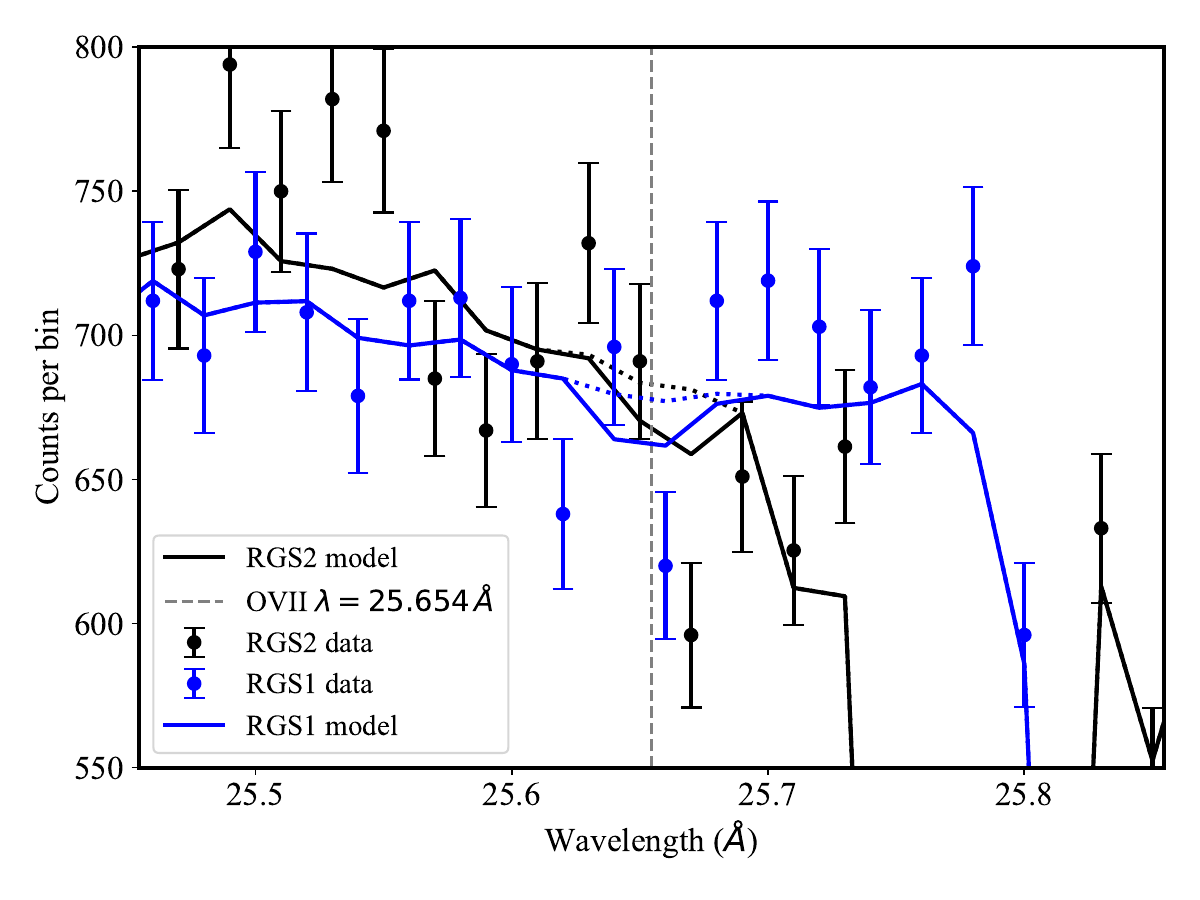}
    \caption{Portions of the \xmm\ spectra of \es. (Left) Region $\pm$1~\AA\ around the 
    location of a putative \ovii\ at the redshift of $z=0.1876$; (b) Zoom-in to illustrate
    the scatter of the data and the depth of the possible absorption.  Note that the best-fit curves do not change upon use of the \cminsys\ for goodness-of-fit estimation.}
    \label{fig:es}
\end{figure}

The methods reviewed in this paper are now discussed and illustrated with the \xmm\ 
spectral observations 
of quasar \es\ from 
\cite{spence2023}. These are the same data that were also described and analyzed in \cite{spence2024} and \cite{bonamente2025c}, to which we refer for details of the data and methods of analysis. The relevant 
statistics are summarized in Table~\ref{tab:es}, and portions
of the relevant spectral data are shown in Figure~\ref{fig:es}. 
The \xmm\ 
 instrument is composed of two independent Reflection Grating Spectrometers detectors, named RGS1 and RGS2, that were both used in the analysis with a joint fit. The RGS data have several regions of reduced efficiency that appear as sudden drops in the 
counts; such regions were excised from the analysis. The error bars in the data are the square 
root of the number of detected counts, which is the standard deviation of a Poisson distribution.
It is to be noted that for the Poisson regression the fit statistic is \eqref{eq:cmin}, which does not make explicit reference to these errors (unlike the Gaussian $\chi^2$ regression). It is visually clear that several data points deviate, possibly in 
a systematic way, from the best-fit model. The methods described in Sec.~\ref{sec:review} are used to estimate
the level of systematics in these spectra and to illustrate how the goodness of fit is 
affected by these systematics.

The model for the broad-band spectra in Fig.~\ref{fig:es} is a spline with break points every \AA\ in an effort to provide the best possible model for the continuum, resulting in a large number of free parameters. Even so, the \cmin\ statistic 
was substantially larger than the number of degrees of freedom (\cmin=1861.8 for 1478 degrees of freedom). According to Eq.~\eqref{eq:sigmaIntEst}, we estimated an average level 
of systematic of $1.8\pm0.2$~\%, which is in fact consistent with the known level 
of systematic uncertainties in the calibration of these instruments, which is estimated
at the few percent level \citep{spence2023}.

A different type of analysis would consist of \emph{assuming} a fixed level of systematics,
say $f_i=0.01$ or $0.05$ ($1$~\% or $5$~\%), and then estimate the distribution of \cminsys\ to determine
whether the measured value of the statistic is consistent with its null-hypothesis
distribution. 
For illustration, a 2\% level of systematics leads to a bias parameter of $\hat{\mu}_C=452.9$,
which indicates that the mean of the \cminsys\ distribution has increased by that amount
relative to the number of degrees of freedom ($\nu=1478$), which is the mean of a $\chi^2_{\nu}$ 
distribution. The overdispersion parameter
is approximately 4 times as large, and as a result the distribution of \cminsys\ becomes
the convolution of the usual $\chi^2_{\nu}$ (which in this regime is approximately
normal with same mean and variance) with a normal distribution $N(\hat{\mu}_C,\hat{\sigma}^2_C)$. It is then clear that the measured value of \cmin=1861.8 becomes perfectly consistent
with the parent distribution; specifically, we find a $p$-value of 0.83. The result is that this level of
systematics explain the measured fit statistic.

In addition to the baseline spline model for
the continuum, a narrow-line model with one additional degree of freedom (the depth of the
line) was added, and thus we have $k=1$ additional parameter in the full model relative to the baseline model. 
With this simple model of the line, the location and width of the line
were held fixed.
For the significance of the nested absorption-line component, in the absence of
systematics a value of $\Delta C=6.6$ falls at the $p=0.010$ value or 99\% level
of significance, based on a $\chi^2_1$ distribution. Such regression would
therefore indicate a `detection' of the line at the 99\% level of confidence, which
some may view as `significant'. 
However, systematics will
lower this level of significance, since the parent distribution of \DeltaCSys\ becomes a randomized chi-squared distribution that has significantly broader tails. Table~1 of \cite{bonamente2025b} provides the critical values of such distribution that can be used to determine
the `corrected' level of significance.~\footnote{The author's \texttt{GitHub} page contains
\texttt{python} software to calculate these distributions.} 
For example, a 2\% level of systematics results in
a Bessel distribution $K(\alpha)$ with a parameter $\alpha= 2 f \sqrt{\mu}$, where $\mu \simeq 700$  is the typical number of counts in the spectrum, for $\alpha \simeq 1.1$ and a corrected
$p$-value of $p=0.012$, which is a negligible correction. However, a  5\% level of systematics
would result in a significantly larger $p=0.036$, which should cast the reality of this line into reasonable doubt \cite{bonamente2025b}.  This example
illustrates the importance of estimating and accounting for systematics for the detection
of faint features.

In summary, the new methods of analysis for Poisson regression reviewed in this paper
provide analytic means to estimate the level of systematic errors in count data, and to
perform an accurate goodness-of-fit assessment, in a variety of data regimes from low-count
to large-count data.  There are several directions for this new models of systematic errors. First is the promotion of its use as a default method for high-energy astrophysics software such as \texttt{XSPEC}, \texttt{SPEX} or \texttt{Sherpa}, by implementation of these analytic routines in the relevant software tools. Second is the continuation of theoretical investigation of certain aspects of the theory that remain unresolved. For example, in \cite{bonamente2025b} we showed that the statistic in Eq.~\ref{eq:DeltaCsys} is composed of two terms that may in fact be correlated, leading to unresolved issues in its distribution. Finally, the relative systematic error can be modeled with more thoughtful parameterizations that the uniform model
presented in Eq.~\eqref{eq:f}, if the applications at hand require such heteroskedastic approach to systematic errors.


\begin{table}[]
    \centering
    \begin{tabular}{l|cl}
    \hline
    \hline
     \multicolumn{3}{c}{Broad-band (13-33\AA) fit for overall goodness of fit}     \\[5pt]
      & & Notes\\[5pt]
     \hline \\
      \cmin\ fit statistic   &  1861.8 & using spline model\\
      d.o.f.    & 1478 & $N-m$ with $N=1526$ and $m=48$\\
      nominal $p$ value & 0.00 & unacceptable fit \\
      $\hat{f}$ (est. systematic error) & $0.018 \pm 0.002$ & estimated from \eqref{eq:sigmaIntEst}\\
      $p$ value for 2\% systematics & 0.83 & acceptable fit (using \cminsys\ distr) \\[5pt]
      \hline 
      \multicolumn{3}{c}{Narrow-band fit for additional absorption-line model component} \\[5pt]
       & & Notes\\[5pt]
      \hline\\
      \cmin\ fit statistic & 103.5 & using a power-law model\\
      d.o.f & 87 & (not needed for absorption line significance)\\
      $\Delta C$ statistic & 6.6 & $k=1$ additional parameter in nested component \\
      nominal $p$ value & 0.010 & nominal detection at 99\% confidence\\
      $p$ value for 2\% systematics & 0.012 & small correction (from distribution \eqref{eq:DeltaCsys})\\
      $p$ value for 5\% systematics & 0.036 & large correction (same distribution) \\
      \hline
      \hline
    \end{tabular}
    \caption{Statistics for the regression  of the \es\ \xmm\ data  using the RGS1 and RGS2 detectors. The narrow-band fit was $\pm1$~\AA\ around a putative \ovii\ ($\lambda_0=21.602$~\AA) absorption line at $z=0.1876$.}
    \label{tab:es}
\end{table}

\newpage
\section*{Conflict of Interest Statement}

The authors declare that the research was conducted in the absence of any commercial or financial relationships that could be construed as a potential conflict of interest.

\section*{Author Contributions}

MB is responsible for all aspects of this publication.

\section*{Funding}
This work was supported by NSF grant 2505174 `\emph{Partnership for Advancing Data
Science Excellence and Education in Southern EPSCoR
States}' awarded to the University of Alabama in Huntsville.



\section*{Data Availability Statement}
Software to reproduce results of this paper are provided at the author's \texttt{GitHub} page.

\bibliographystyle{Frontiers-Harvard} 
\bibliography{max}


\end{document}

%% file: defs.tex
\newcommand\Var{\text{Var}} 
\newcommand\E{\text{E}} 

\def\Likelihood{\mathcal{L}}

\def\DeltaC{$\Delta {C}$}
\def\DeltaCSysEq{\Delta {C}_{\mathrm{sys}}}
\def\DeltaCSys{$\DeltaCSysEq$}

\def\ovii{O~VII}
\def\1es{1ES~1553+113}

\def\cmin{$C_{\mathrm{min}}$}
\def\cmineq{C_{\mathrm{min}}}

\def\cminsys{$C_{\mathrm{min,sys}}$}
\def\cminsyseq{C_{\mathrm{min,sys}}}

\def\DeltaC{$\Delta C$}
\renewcommand\Var{\text{Var}}

\renewcommand\E{\text{E}}

\def\Poiss{\mathrm{Poisson}}

\def\cminValEq{c_{\mathrm{min}}}
\def\cminVal{$\cminValEq$}
\def\xmm{\emph{XMM--Newton}}
\def\es{1ES~1553+113}

\def\sigmaIntiEq{\sigma_{\mathrm{int},i}}